\documentclass{aa}  
\usepackage{graphicx}
\usepackage{txfonts}
\begin{document} 

   \title{Flat-spectrum radio quasars as high-energy neutrino sources}
   \author{A. Moretti\inst{1}
          \and
          A. Caccianiga \inst{1} }
   \institute{INAF - Osservatorio Astronomico di Brera\\
              \email{alberto.moretti@inaf.it}\\
             \email{alessandro.caccianiga@inaf.it} }
   \date{Received ...; accepted ...}
\abstract{The astrophysical sources responsible for the production  of high-energy  neutrinos remain largely uncertain. The strongest associations suggest a correlation between  neutrinos and active galactic nuclei.
However, it is still unclear which specific regions and mechanisms of the accreting supermassive black hole are responsible for their production. In this paper we investigate the correlation between the positions of IceCat-1 neutrino events and a large, optically selected quasar catalogue 
extracted from the Sloan Digital Sky Survey. Within this sample, we distinguish radio-quiet quasars from flat-spectrum radio quasars (FSRQs)
based on radio emission data from the Cosmic Lens All Sky Survey (CLASS) catalogue. While all the associations between neutrino events and radio-quiet quasars are consistent with being random matches, FSRQs exhibit a moderately significant correlation ($\sim$2.7$\sigma$) with neutrino positions. 
Additionally, we observe that the distribution of minimum distances between neutrino events and FSRQs differs significantly for events at declinations above and below 20$^\circ$.
In particular, using the Kolmogorov-Smirnov test, we find that the high-declination event distribution deviates strongly (4$\sigma$) from a random distribution. 
We interpret all these results as an indication that a large fraction of the neutrino events  ($>$60\%) observed by IceCube could be produced by the FSRQs and that the emission mechanism is likely related to the relativistic jets rather than the radio-quiet component of these sources, such as the accretion disk or corona. 
}
   \maketitle
%
\section{Introduction}
\label{sec:intro}
Active galactic nuclei (AGNs) are commonly considered promising candidates for the emission of the high-energy astrophysical neutrinos that have routinely been detected by the IceCube experiment since 2013 (\citealt{Aartsen2013} for a review). Indeed, AGNs can be powerful accelerators of protons, either through relativistic jets or through the accretion disk, hot corona. These particles can then interact with the intense radiation fields present around the active nucleus, thereby producing cascades of particles, including energetic neutrinos. The emission from relativistic jets pointing towards Earth in particular can be very strong due to relativistic boosting (beaming). This is the reason why these `oriented' sources, called blazars, have been extensively investigated in the last $\sim$10 years, to assess their possible relevance for neutrino astrophysics (\citealt{Dermer2014a,Tavecchio2014,Tavecchio2015,Giommi2020a,Giommi2020b,Giommi2021,Plavin2020,Plavin2023,Buson2019,Buson2022,Buson2023,Bellenghi2023}).

Blazars can be observed in a wide variety of `flavours' depending on the shape of the spectral energy distribution and/or the optical properties. In particular, blazars in which the synchrotron emission peaks at low ($<$10$^{14}$ Hz), intermediate (10$^{14}$-10$^{15}$ Hz), or high ($>$10$^{15}$ Hz) frequencies are called low-, intermediate-, and high-frequency peak sources, respectively (\citealt{Padovani1995,Abdo2010}). 
Based on their optical properties, blazars with featureless spectra are instead dubbed BL Lacertae (BL Lac) objects; they are called flat spectrum radio quasars (FSRQs) when strong and broad emission lines, similar to those usually found in non-jetted AGNs, are observed. The distinction between BL Lac and FSRQs could be relevant for neutrino physics since the very existence of a broad-line region (BLR) and a `standard' accretion disk is directly connected to the presence of the radiation field necessary for an efficient neutrino emission (e.g. \citealt{Dermer2014, Murase2014}).
For this reason, FSRQs were originally considered the most promising neutrino emitters, although BL Lacs could also be viable candidates if velocity structures are present in the jets (e.g. \citealt{Tavecchio2014, Tavecchio2015}). 
\begin{figure*}
\begin{center}
\includegraphics[width=13cm]{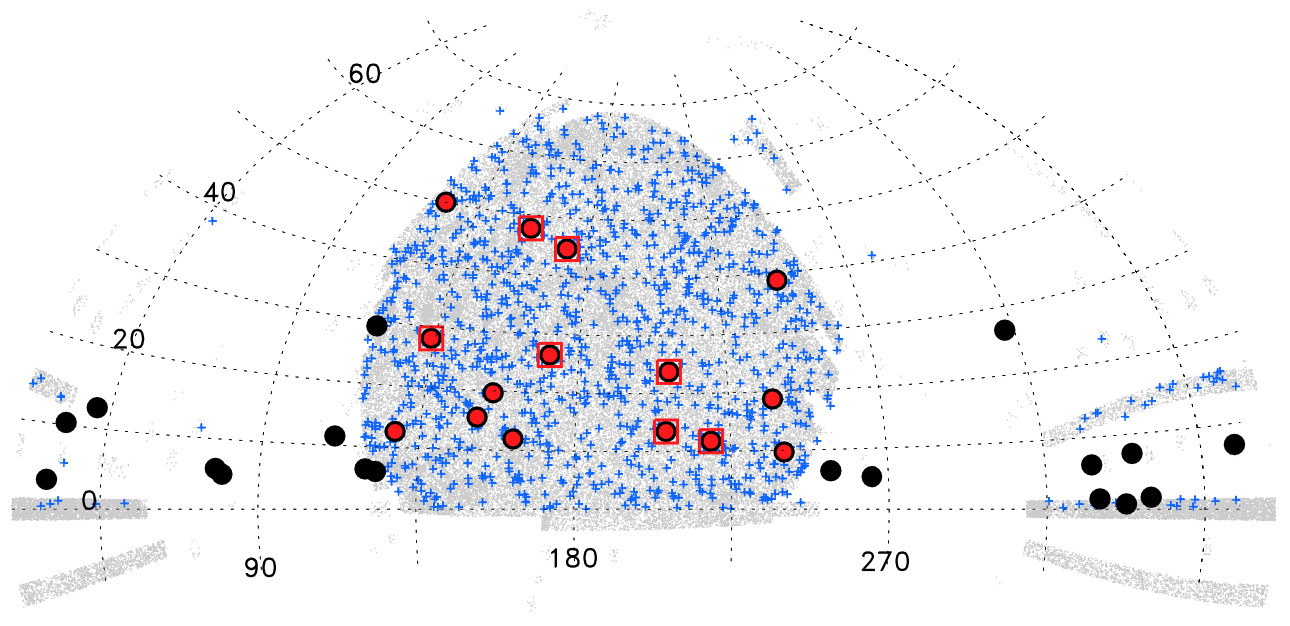}
\caption{Sky distribution of the data. The small grey dots represent the positions of 105,783 quasars from the S11 catalogue.
The small blue crosses indicate the 1,218 CLASH FSRQs. Black circles mark the 33 IceCAT-1 events in the northern sky with the required positional accuracy. Circles with a red centre represent the 15 neutrinos located within our area of interest uniformly covered by the CLASH objects. A red square around a circle indicates that a CLASH FSRQ is included in the nominal error region. 
\label{fig:map}}
\end{center}
\end{figure*}

Interestingly, the first firm association  ($>$3$\sigma$) between a neutrino event and an extragalactic source, based on the detection of flares temporarily coincident with a neutrino's arrival and then confirmed by an excess of events in the archival IceCube data,
was the source TXS~0506+056 (\citealt{Aartsen2018a,Aartsen2018b, Aartsen2020}), an intermediate-frequency peak source
that was originally classified as a BL Lac object. 
However, a detailed analysis of its optical spectrum revealed the presence of a `standard' accretion disk emission (as in FSRQs), whose light was probably swamped by the strong non-thermal continuum coming from the oriented relativistic jet (\citealt{Padovani2019}). This observation sets TXS~0506+056 apart from `true' BL Lac objects that genuinely lack of an efficient accretion disk and a BLR, probably due to the low accretion rates of their central super massive black holes (SMBHs; e.g. see \citealt{Hardcastle2020}), and suggests an emission mechanism that is similar to that considered for FSRQs \citep{Keivani2018}.

The search for positional coincidences between neutrino events (or excesses of events) and extragalactic sources is hampered by the large positional uncertainties of the events.\ These uncertainties make the number of chance coincidences very high, even for relatively rare objects like blazars. 

Hints of positional correlations between neutrino events and BL Lac objects have been found by \citet{Padovani2016} and \citet{Giommi2020b}.  Other works have suggested a possible association between IceCube events and the most radio-luminous blazars ($>$150~mJy), including both FSRQs and BL Lacs \citealt{Plavin2020, Plavin2021, Plavin2023, Buson2022, Buson2023}.

Even if relativistic jets are natural candidates for neutrino emission, the recent association of a radio-quiet (RQ) AGN (NGC1068) with neutrino events (\citealt{Abbasi2022a}) suggests that other mechanisms are at work. For instance, \citet{Abbasi2022b} have shown that the acceleration of cosmic rays responsible for the neutrino emission can occur in the accretion disk region of AGNs. 
The multi-wavelength analysis of NGC 1068  conducted within a p-$\gamma$ scenario indicates that the most plausible location for the neutrino source is in close proximity to the SMBH, which is embedded in an extremely photon-dense environment \citep{Murase2022, Padovani2024}. 
This finding suggests the AGN corona as the likely origin, while star-forming regions, the jet, and the outflow are disfavoured.
\citet{Yang2025} find that even in the case of TXS~0506+056, the accretion flow might explain both the long-term steady neutrino emission and the outburst, for which a super Eddington accretion is required. 
For this reason, the simple positional match of a sample of blazars with neutrino events, even when statistically significant, does not necessarily imply that the emission is due to the relativistic jets, since some of these sources (the FSRQs) also have a powerful accretion disk similar to that observed in RQ quasars.

In this paper we test the hypothesis that FSRQs are responsible, at least in part, for the astrophysical neutrino events detected by IceCube. At the same time, we aim to establish whether this is really due to the presence of an oriented jet or, instead, can be associated with the RQ component of a FSRQ, i.e. the accretion disk and corona. To this end, we started from an optically selected sample of AGNs (that includes both radio-loud and RQ objects), and from this we extracted both the FSRQ and the 
RQ quasar samples. From an analysis of their correlation with neutrino events, we assessed not only the association of neutrino production with quasars, but also the possible origin site of the neutrinos (jets or disk-corona accretion).

In Sect.~2 we present the quasar data sample and the list of IceCube events that we considered in the analysis.
In Sect.~3 we describe the positional quasar--neutrino cross-match and present the results for both FSRQs and RQ quasars. 
In Sect.~4 we analyse the distribution of the minimum distance between neutrinos and quasars.  
In Sect.~5 the most likely neutrino emitter candidates are presented individually, and  
in Sect.~6, we discuss the purity of the sample together with the estimated fraction of high-energy neutrino events 
produced by FSRQs.
Finally, in Sect.~7 we put the results in the context of recent findings presented in the literature. 
\begin{table*}
\scriptsize 
 \centering
 \caption{Neutrino events considered in our analysis.}
 \label{tab:matches0}
 \begin{tabular}{l c c c c c | c c c}
  \hline
         Event &  ra   & dec & energy   &    type  & sign. & CLASH & dist  & flux \\ 
               & [deg] & [deg]& [Tev]   &          &       &       & [deg] &  [mJy\@5Ghz]\\  
  \hline
    IC110929A & 121.450$_{-1.29}^{+1.34}$  & 50.040$_{-0.15}^{+0.24}$ &   158.0 &    gfu-bronze & 0.521 & -- & -- & --\\
    IC111012A & 172.130$_{-1.39}^{+1.40}$  & 44.700$_{-0.45}^{+0.79}$ &   115.0 &    gfu-bronze & 0.434 & GB6J1126+451 &  0.6318 &    360.0\\
    IC120301A & 237.960$_{-0.62}^{+0.53}$  & 18.760$_{-0.51}^{+0.47}$ &   433.0 &      gfu-gold & 0.825 & -- & -- & --\\
 IC120523A(*) & 171.080$_{-1.41}^{+0.66}$  & 26.440$_{-0.37}^{+0.46}$ &   213.0 &      ehe-gold & 0.525 & GB6J1125+263 &  0.2983 &     38.0\\
              &                            &                          &         &               &       & GB6J1125+261 &  0.4431 &   1176.0\\
    IC150904A & 133.770$_{-0.88}^{+0.53}$  & 28.080$_{-0.55}^{+0.51}$ &   302.0 &      gfu-gold & 0.741 & GB6J0853+281 &  0.4208 &    109.0\\
    IC160331A & 151.220$_{-0.66}^{+0.66}$  & 15.480$_{-0.73}^{+0.66}$ &   492.0 &      gfu-gold & 0.851 & -- & -- & --\\
    IC160427A & 240.290$_{-0.48}^{+0.44}$  &  9.710$_{-0.42}^{+0.57}$ &    85.0 &   hese-bronze & 0.451 & -- & -- & --\\
    IC180125A & 207.510$_{-0.57}^{+1.01}$  & 23.770$_{-0.57}^{+0.57}$ &   110.0 &    gfu-bronze & 0.362 & GB6J1350+241 &  0.4381 &    154.0\\
              &                            &                          &         &               &       & GB6J1350+233 &  0.2919 &    141.0\\
    IC190201A & 245.080$_{-0.88}^{+0.75}$  & 38.780$_{-0.67}^{+0.77}$ &   163.0 &    gfu-bronze & 0.533 & -- & -- & --\\
    IC190223A & 155.210$_{-0.66}^{+0.70}$  & 19.670$_{-0.44}^{+0.28}$ &   168.0 &    gfu-bronze & 0.512 & -- & -- & --\\
    IC190413A & 219.330$_{-1.32}^{+0.70}$  & 11.720$_{-0.72}^{+0.72}$ &   107.0 &    gfu-bronze & 0.292 & GB6J1439+111 &  0.6271 &    133.0\\
    IC190515A & 127.880$_{-0.83}^{+0.79}$  & 12.600$_{-0.46}^{+0.50}$ &   457.0 &      gfu-gold & 0.816 & -- & -- & --\\
    IC200620A & 162.110$_{-0.92}^{+0.62}$  & 11.950$_{-0.46}^{+0.61}$ &   114.0 &    gfu-bronze & 0.325 & -- & -- & --\\
    IC200806A & 157.250$_{-0.87}^{+1.17}$  & 47.750$_{-0.59}^{+0.64}$ &   107.0 &    gfu-bronze & 0.397 & GB6J1027+480 &  0.4260 &    242.0\\
    IC201222A & 206.370$_{-0.75}^{+0.88}$  & 13.440$_{-0.34}^{+0.54}$ &   186.0 &      gfu-gold & 0.534 & GB6J1348+133 &  0.7439 &     41.0\\
\hline
 \end{tabular}
 \vspace{0.2cm}

\tablefoot{* In the Ice-Cat-1 catalogue, two events named IC120523A are present; the one listed here is different from the one associated with 3C454.3 by \cite{Abbasi2023b}.}. 
\end{table*}

Throughout the paper we assume a flat $\Lambda$ cold dark matter cosmology with H$_0$=71 km 
s$^{-1}$ Mpc$^{-1}$, $\Omega_{\Lambda}$=0.7, and $\Omega_{M}$=0.3.  Spectral 
indices are given assuming S$_{\nu}\propto\nu^{-\alpha}$.

\section{Quasar and neutrino catalogues}
\subsection{The quasar samples}
We started with the sample of type~1 AGNs presented in \citet[hereafter the S11 sample]{Shen2011}, which is based on the data from the Sloan Digital Sky Survey (SDSS).  The sample includes 105,783 spectroscopically confirmed type~1 AGNs, mostly in the northern hemisphere, and contains several pieces of information on the central engine, such as the mass of the SMBH and its accretion rate, which can be used to further investigate potential neutrino emitters. 
For our analysis we only considered the sky area within 120$^{\circ} <$ R.A. $<$260$^{\circ}$ and 0$^{\circ}$ Dec. 60$^{\circ}$, where the coverage is uniform (Fig.~\ref{fig:map}); this area includes a total of 89,437 quasars.

For the radio information we used the Cosmic Lens All Sky Survey (CLASS; \citealt{Myers2003,Browne2003}), which homogeneously covers all the extragalactic northern sky. The CLASS catalogue contains $\sim$11,000 flat spectrum radio sources selected at 5~GHz down to a well-defined flux density limit (30~mJy). Since blazars are flat spectrum radio sources, this survey represents the best starting point for our selection. The survey is also deep enough to select most of the blazars with an optical magnitude brighter than 20 (which roughly corresponds to the limit of the S11 sample). Thanks to the excellent positional accuracy of the survey ($<$1\arcsec) the optical/radio association can be performed with high reliability (Fig.~\ref{fig:map}).
The cross-correlation between the S11 sample and CLASS produces 1218 sources (1122 in the region of interest), using 1\arcsec\ positional tolerance. 
We call this sample CLASH, and throughout the rest of the paper, unless otherwise specified,
flux densities are measured at 5 GHz.

For comparison, we also built a sample of RQ quasars, defined as those sources from S11 not detected in CLASS and located more than 5$\arcsec$ from any source in the FIRST catalogue. FIRST is a radio survey at 1.4 GHz with flux density limit of $\sim$1 mJy, significantly deeper than CLASS and with a good positional accuracy ($\sim$ 2$\arcsec$). Given the typical magnitudes of the \citet{Shen2011} quasars, the non-detection of a source in the FIRST survey implies a radio-loudness\footnote{The radio-loudness is defined as the ratio between the radio and the optical monochromatic luminosities (per unit of frequency) computed, respectively, at 5~GHz and at 4400\AA\  (\citealt{Kellermann1989})} below 10 i.e. in the typical RQ regime (\citealt{Kellermann1989}).
Therefore, the RQ quasar sample should mostly contain non-jetted objects.  This control sample comprises 73,339 RQ objects.

\begin{figure}
\includegraphics[width=\columnwidth]{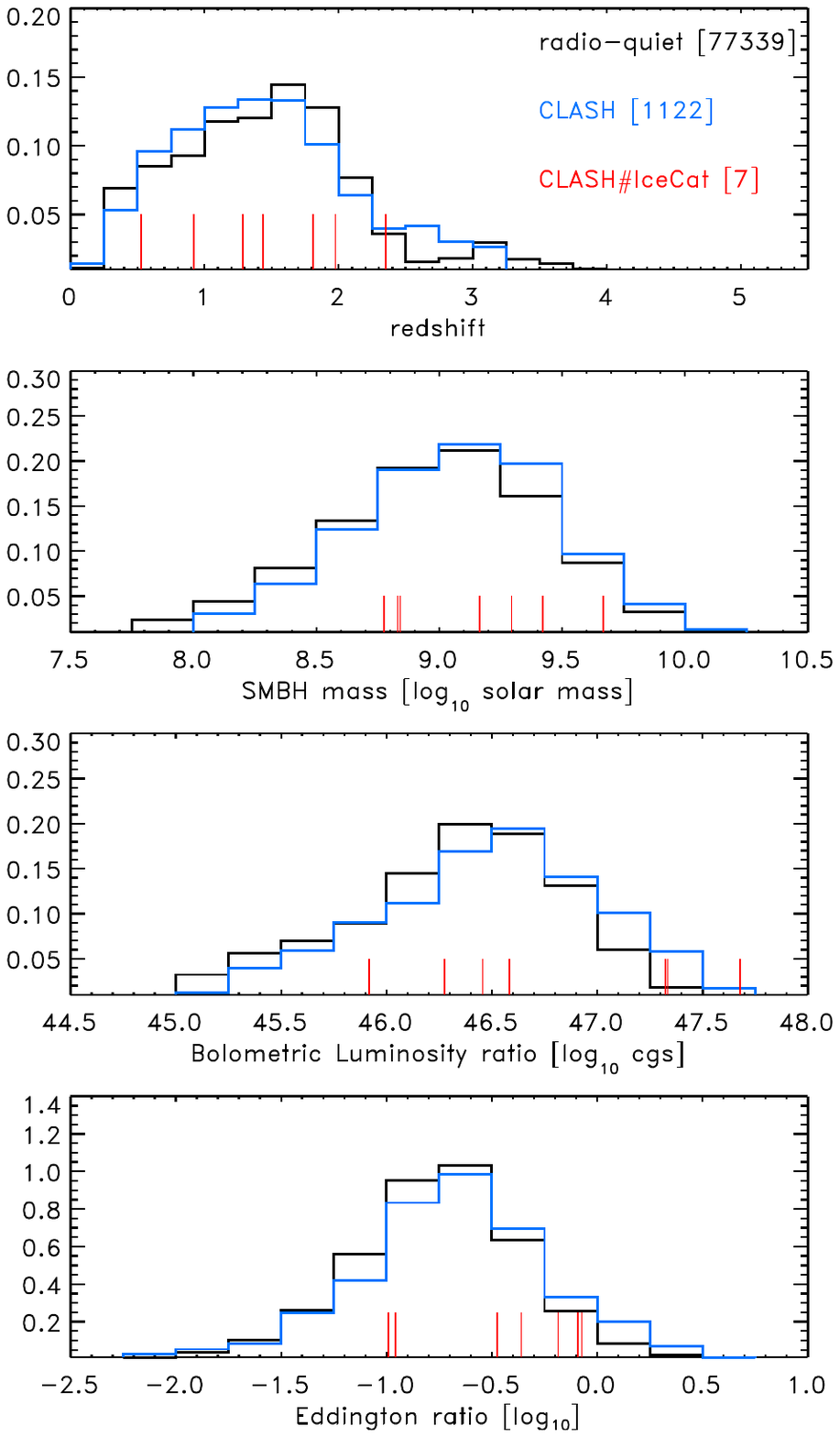}  
\caption{Comparison of the properties of CLASH and RQ quasars. The distributions of the redshift, mass, bolometric luminosity, and Eddington ratio of the RQ sample (black) and CLASH blazars (blue) are shown. The values of the possible neutrino emitters are given in red (Table~\ref{tab:matches1}).
\label{fig:histprop}}
\end{figure}

As shown in Fig.~\ref{fig:histprop}, the two populations are indistinguishable in terms of redshift, mass, bolometric luminosity and Eddington ratio. The only difference is the presence of a strong radio emission, which we assume is produced by a relativistic jet oriented towards our line of sight.

\subsection{The IceCube neutrino events}\label{sect:neut}
For our study we used the IceCube Event Catalog of Alert Tracks (IceCat-1; \citealt{Abbasi2023a}), which includes 267 non-vetoed track-like neutrino events likely produced by astrophysical sources between 2011 and 2021.
Since the positional uncertainty is extremely large in a significant fraction of the events,   
we considered only those events with the best positional accuracy: 
\begin{equation}
(ERR_{RA}^- + ERR_{RA}^+)cos(DEC)<2.0 ^{\circ}
\end{equation}
\begin{equation}
ERR_{DEC}^- + ERR_{DEC}^+<2.0 ^{\circ}.
\end{equation}

\begin{figure}[t]
\includegraphics[width=\columnwidth]{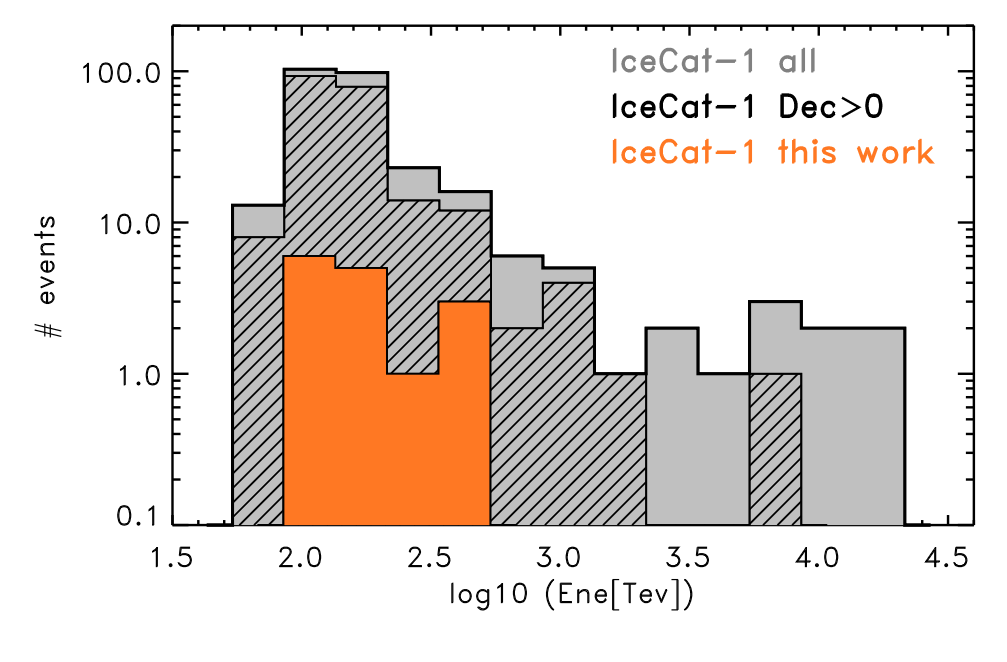}  
\caption{ Energy distribution of the 15 selected events (red) compared to the energy distributions of all  267 Icecat-1  events (grey) and those observed in the northern sky (hatched grey).
\label{fig:histene}}
\end{figure}

We stress that these requirements were frozen at the beginning of the analysis, and no optimization has been performed. 
In the region of interest defined above (Fig.~\ref{fig:map}), these constraints produce a list of 15 events (Table ~\ref{tab:matches0}),
with  IC121103A, IC190503, and IC200117A excluded because too close to the border.
Following the approach of \citet{Plavin2023}, also discussed in \citet{Abbasi2023b}, the 90\% error region is assumed to be encircled by the 4 elliptical segments with the RA and Dec. errors as semi-axes. 
The energy distribution of the 15 events  is shown in  Fig.~\ref{fig:histene}: the selection on the positional accuracy includes events belonging to the bulk of the population detected in the northern sky with energies in the 100-500 TeV interval (Table \ref{tab:matches0}).

\section{Quasar--neutrino associations}
\subsection{CLASH FSRQ--neutrino association}
Considering the list of 15 IceCat-1 events and the 1122 CLASH FSRQ we find that 9 CLASH sources fall within the error boxes of 7 IceCube events (Table~\ref{tab:matches0}).  
Varying the flux density limit of the CLASH sample, the number of objects varies from 1122 with a limit of 30 mJy to 270, when the flux density limit is 200 mJy. 
The number of neutrino events, with at least one CLASH quasar within the 90\% error region varies from 7 to 3, while the number of CLASH included 
in the neutrino  90\% confidence error region varies from 9 to 3 (Table~\ref{tab:montec}). 

Multiple matches clearly indicate that the probability of chance coincidence is not negligible.
We estimated the expected number of chance coincidences by randomly shifting quasar positions within a circular annulus of radius 3–10$^{\circ}$ around their original locations. This procedure was repeated 10$^5$ times for different radio flux density limits of the FSRQ sample. 
The statistical significance of the observed correlation (p-value) was calculated by counting the realizations in which the number of neutrinos including at least one CLASH FSRQ exceeded that observed in the actual dataset. The output of this Monte Carlo simulation at different radio flux density limits are reported in Fig.~\ref{fig:p_vs_flux} and Table~\ref{tab:montec}. 

These values show a minimum located at a flux density limit of $\sim$100 mJy (Fig.~\ref{fig:p_vs_flux}).
At this flux density limit, the number of CLASH FSRQs is 497, with 7 of them in the error box of 6 neutrino events. 
The average number of associations of an equal number of random positions with the 15 neutrino error box is $\sim$ 1.3. 
We find that in 52 cases out of 1e5 an equal or larger number neutrino have a CLASH in the error box. 
Considering the number of different trials involved in this analysis (6 with different flux density limits), we estimate that the final probability  corresponds to a significance of $\sim$2.7$\sigma$.
We obtain identical results when repeating the simulation by randomly shifting the positions of the neutrinos instead of the quasars.
\begin{table}[h!]
\scriptsize 
 \centering
 \caption{p-values of the neutrino-FSRQ CLASH associations at different flux density limits.}
  \label{tab:montec}
 \begin{tabular}{l c |c c c| c c}
  \hline
f$_{min}$ & Quasars & \#n &  $<$\#n$>$ &    p-val    & KS  & KS ($\delta>$20$^\circ$)   \\ 
  \hline
  30 &     1122   &    7   &    2.6  &   6.3e-03  &  2.8e-02  & 8.2e-05   \\
  75 &      614   &    6   &    1.4  &   1.7e-03  &  6.5e-03  & 7.8e-06  \\
 100 &      497   &    6   &    1.2  &   5.3e-04  &  3.3e-03  & 4.8e-06   \\
 125 &      402   &    5   &    0.9  &   1.5e-03  &  2.4e-03  & 9.4e-06  \\
 175 &      300   &    3   &    0.7  &   3.0e-02  &  8.4e-03  & 5.3e-06  \\
 200 &      270   &    3   &    0.6  &   2.3e-02  &  2.9e-02  & 1.4e-04  \\ 
\hline
\end{tabular}

\tablefoot{
(i)    Flux density limit of the FSRQ sample (mJy at 5GHz);   
 (ii)   number of CLASH objects;
 (iii)  neutrinos with at least 1 CLASH in the error box;
 (iv)    mean of neutrino error boxes containing at least one CLASH source in the Monte Carlo random samples;
 (v)   probability of having simulated dataset with equal or more neutrino error boxes containing at least 1 CLASH  (p-value);
 (v,vi) probabilities that the minimum distance distribution come from a random position sample}

\end{table}

\subsection{RQ quasar--neutrino association}

Among the 77339 RQ quasars, 178 fall within the 15 neutrino error regions.
Since these numbers do not allow the assessment of any possible significant
correlation, we followed two different approaches.
First, we compared the RQ quasars to the FSRQ sample with similar statistic, using the 77339 RQ sources to construct 10,000 different random samples of 1,122 RQ objects each.
For each sample, we repeated the same analysis performed on the CLASH FSRQs, extracting subsamples of the same size as the CLASH sample at different flux density limits, ranging from 1122 to 270 objects.
For each of these samples, we counted the matches with the 15 neutrino positions and estimated the expected number of chance coincidences again by randomly shifting the positions.
We then repeated the same analysis on the 1,122 most massive and the 1,122 most luminous quasars, regardless of their radio flux.
\begin{figure}[ht!]
\includegraphics[width=\columnwidth]{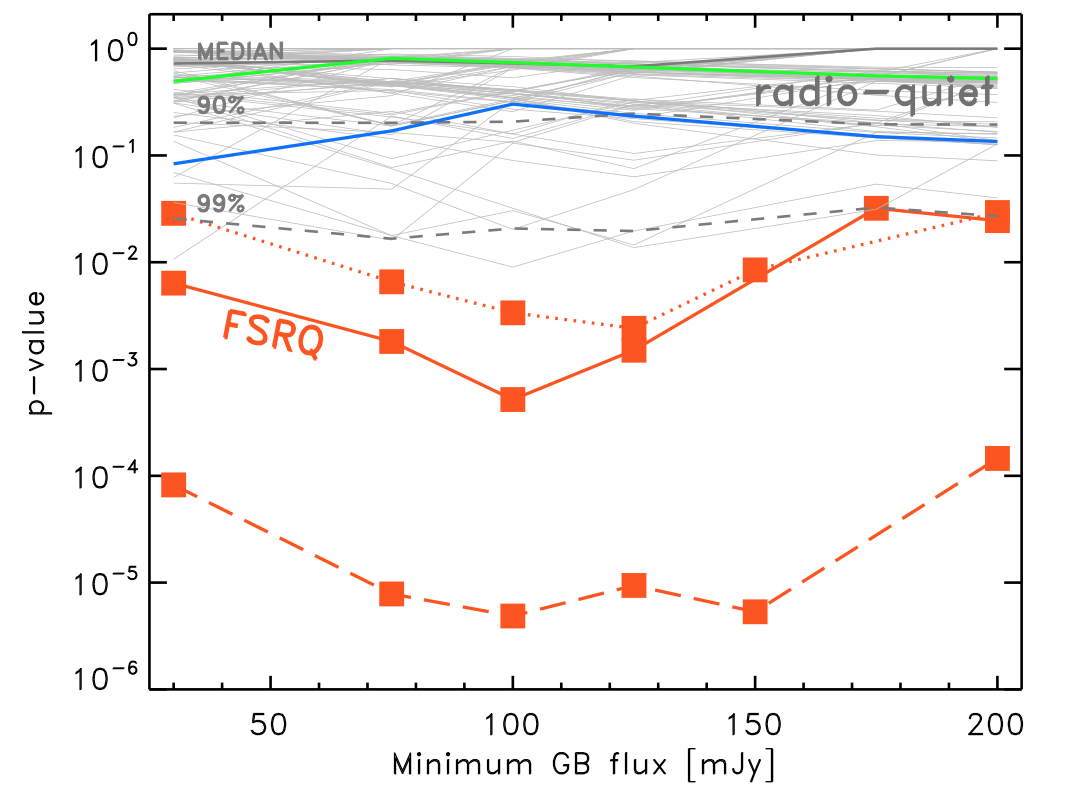}  
\caption{
p-values for the neutrino--CLASH associations (red) and neutrino--S11-RQ quasar associations (grey).
The solid red line, plotted for different flux density limits, represents the probability of randomly finding an equal or greater number of neutrino events with at least one CLASH FSRQ within the error box.
The dotted and dashed lines indicate the probabilities that the distribution of minimum distances between neutrinos and CLASH sources arises from a random distribution (KS test) for all 15 events and for the high-declination sample, respectively.
The p-values for 100 RQ samples, of the same size as the FSRQ sample at the corresponding radio flux limit, are displayed in grey.
The median value, along with the 90th and 99th percentiles, is calculated from 10,000 samples, which are not shown for clarity.
We plot the p-values for the most luminous and most massive quasars of the S11 sample in green and blue, respectively.
\label{fig:p_vs_flux}}
\end{figure}
The p-value results are shown in Fig.~\ref{fig:p_vs_flux}.
We find average p-values of $\sim$0.7 with the 99\% of the samples having p-values higher than 0.02, meaning that the number of RQ quasars present in the neutrino error box is fully consistent with what expected from chance coincidence.

The absence of a correlation between RQ quasars and neutrino events,
combined with the correlation found for the FSRQ subsample, strongly suggests
that, if FSRQs are confirmed as neutrino sources, the production mechanism
must be tied to the physics of the jet rather than the accretion flow.

\section{Neutrinos and the closest CLASH FSRQs}
\begin{figure}[t]
\includegraphics[width=\columnwidth]{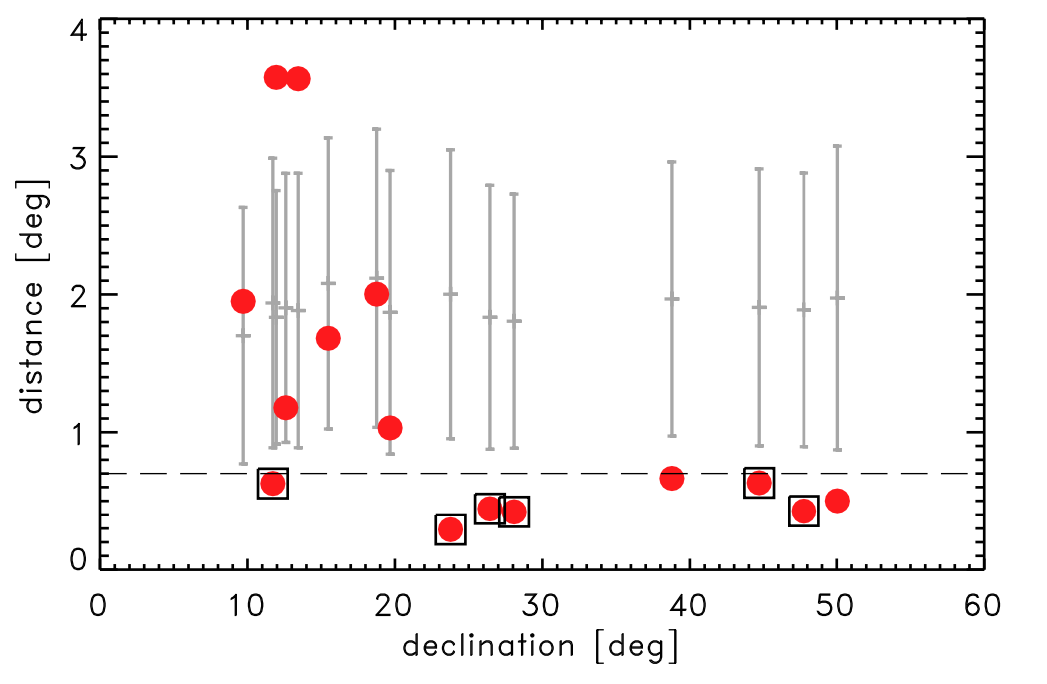}  
\caption{Distances between the 15 neutrino events and the closest CLASH FSRQs  as a function of the declination (red circles). A square around the circle indicates that the CLASH FSRQ is included in the nominal error region. For each event, in grey, we plot the mean and standard deviation of the distribution of the minimum distances of {a quasar sample of the same size with randomized positions.}
\label{fig:decerr}}
\end{figure}

In the previous section we assessed the correlation between neutrinos and CLASH objects by counting the number of neutrino events that include FSRQs within their 90\% confidence error regions.
In this section we present a different approach, analysing the angular distances between each of the 15 considered neutrino events and the nearest CLASH source.
Figure~\ref{fig:decerr} presents the minimum distances between the 15 neutrino positions and CLASH sources as a function of declination, considering only objects with a flux density above 100 mJy (497 sources).
The figure also shows, for comparison, the mean and standard deviation of the distribution of minimum distances obtained from random control samples.  Specifically, we generated 10,000 control samples of 497 elements, randomly extracted from the RQ population, which uniformly covers the same area as the CLASH sample. As an additional check, we produced 10,000 further samples by randomizing the positions within a circular annulus of radius 3–10$^{\circ}$ around the original locations of the CLASH and RQ quasars, finding perfectly consistent numbers.

A clear trend emerges when comparing the subsamples at low and high declinations. At low declinations, the distances between seven out of the eight neutrino events and the nearest CLASH source are consistent
with random expectations. In contrast, at higher declinations, all seven neutrino events have a nearby CLASH bright object within $\lesssim$ 0.7$^\circ$, significantly closer than expected from random distributions.

We applied a Kolmogorov-Smirnov (KS) test to evaluate whether the CLASH-neutrino distance distribution differs significantly from a random distribution.
The results, for both the full sample of 15 neutrino events and the high-declination subsample (7 events), are presented in Fig.~\ref{fig:p_vs_flux}
and summarized in Table~\ref{tab:montec} for different radio flux density thresholds.
While the difference is moderately significant for the full sample, the high-declination subsample shows a much stronger deviation from randomness. The minimum uncorrected p-value at flux limit of 100 mJy is 4$\times$10$^{-6}$, corresponding to a significance of 4.0$\sigma$ after accounting for the six trials (Fig.~\ref{fig:p_vs_flux}).
\begin{figure*}
\begin{center}
\includegraphics[width=17cm]{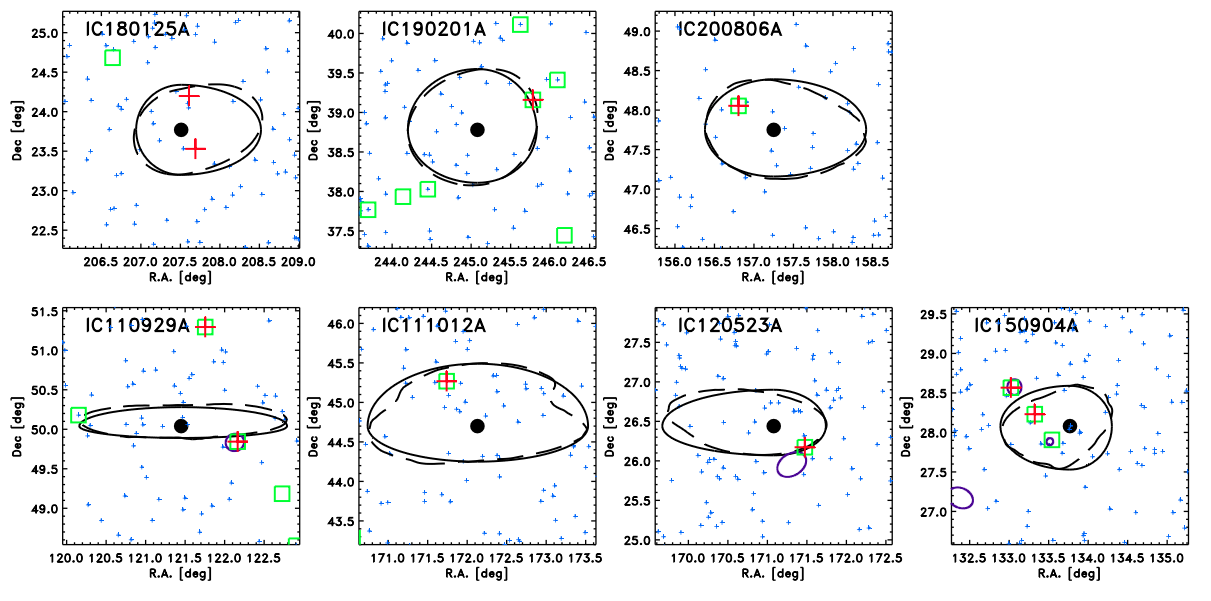}  
\caption{Maps of the seven CLASH objects identified as  possible neutrino sources. Black circles show nominal neutrino positions. Dashed lines show the 90\% confidence error regions, and the solid lines show the elliptical approximation. The CLASH FSRQs are indicated by red crosses. Green squares show BZCat blazars, and purple ellipses the \textit{Fermi} catalogue element 90\% confidence error regions. The small blue crosses are the S11 quasar positions. 
\label{fig:footp}}
\end{center}
\end{figure*}
The dependence of the distance between the closest CLASH-FSRQ and neutrino events on declination may have a technical explanation. The performance of the IceCube telescope, particularly in terms of background levels and effective area, varies significantly with declination (\citealt{Abbasi2023a}), which corresponds to the (opposite of the) neutrino incident direction.
Although we are not in a position to fully assess the event detection error budget, our results suggest that the positional accuracy and purity level of IceCat-1 catalogue events are higher at declinations above 20$^{\circ}$. 

As the number of events with good positional accuracy in IceCat-1 within our region of interest is limited, in order to check this result, we expanded our sample.
We considered the entire CLASS catalogue, regardless of identification with an S11 quasar.
To be homogeneous with the S11 catalogue, we considered only sources with magnitudes of $<$20 in at least one of the SDSS band.
With this limit at flux densities of $>$100 mJy, the CLASS catalogue includes 1790 objects, of which 1052 are classified as FSRQs in the literature; the remaining objects are unclassified.

\begin{figure}[ht!]
\includegraphics[width=\columnwidth]{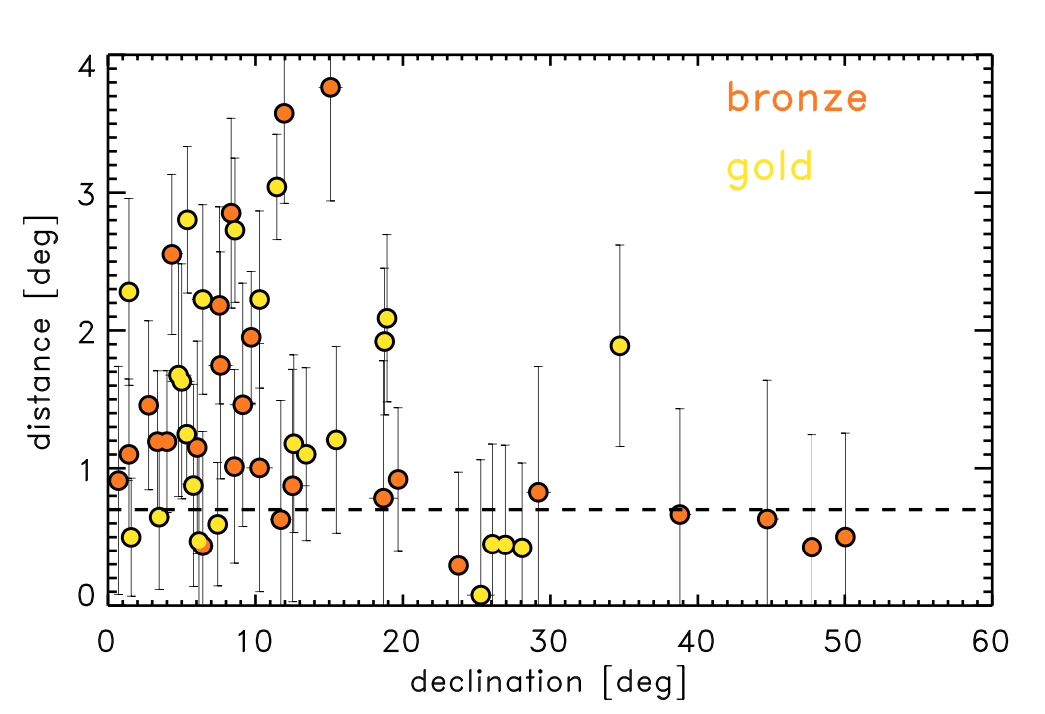}  
\caption{Same as Fig.~\ref{fig:decerr} but for a larger sample of neutrino events. 
The radio counterparts in this case are taken from the CLASS catalogue, regardless of identification with an S11 quasar.
\label{fig:decerrc}}
\end{figure}

This extension allowed us to use all the 33 events in the northern hemisphere with good positional accuracy (as defined earlier). Additionally, we included the 18 neutrino events that meet the same positional accuracy criteria and occurred  from January 2022 to December 2024, as reported in the GCN catalogue\footnote{ https://gcn.gsfc.nasa.gov/amon\_icecube\_gold\_bronze\_events.html}.
The distances between these 51 events and the closest CLASS source are reported in Fig.~\ref{fig:decerrc}.
Among the four objects added to the high-declination sample, three have a minimum distance $\lesssim$0.4$^\circ$, while one is at 1.8$^\circ$. We note that one of the three at a short distance is GB6J0740+28, a CLASH object in the error region of IC200117A, which was excluded from the sample of 15 events because it is too close to the border of our region of interest (Sect.~\ref{sect:neut}).
Therefore, the pattern of high-declination events being closer to a bright radio source seems to also be present in this expanded catalogue. 

\section{High-declination sample}
Focusing on the original sample of 15 events, we find that, among the seven neutrinos detected at declinations above 20$^{\circ}$, five have a bright CLASH FSRQ (flux $>$100 mJy) within their 90\% confidence error region, while all seven have at least one CLASH FSRQ within 0.7$^{\circ}$ (Fig.~\ref{fig:decerr}).
We note that the two high-declination events without a bright CLASH source within their nominal 90\% confidence error region would include such a source if a systematic error of 0.1$^{\circ}$ were (linearly) added to their nominal positional uncertainties (Fig.~\ref{fig:footp} ).

In this section we highlight the main characteristics of these seven CLASH sources.
All the information presented here is derived from public databases and catalogues. 
The seven FSRQs have redshifts in the 0.5-2.3 range and  optical AB magnitudes between 17 and 19. 
X-ray fluxes in the 0.5-2.0 keV band vary between 1 and 5$\times$ 10$^{-13}$ erg s$^{-1}$ cm$^{-2}$, which are typical values for FSRQs with 
radio fluxes $>$ 100 mJy. 
The SMBH masses range between 10$^{8.8}$ and 10$^{9.7}$ solar masses and are accreting at rates estimated between 0.1 and 0.8 in terms of Eddington ratios (Table~\ref{tab:matches1}). We investigated whether these numbers are consistent with the characteristics of the S11 RQ quasar population. The analysis is hampered by the small sample size and no significant difference was detected, with K-S test probabilities exceeding 10\% in all three cases.
 
 From a visual inspection of the spectral energy distribution of these seven objects, determined using the {\it FIRMAMENTO}\footnote{https://firmamento.hosting.nyu.edu/home} tool (\citealt{Giommi2025}), we find that they all have a usual blazar shape, with a synchrotron peak falling at 10$^{13}$-10$^{14}$ Hz, i.e. in the typical range observed in FSRQs.
Only one object listed in Table~\ref{tab:matches1} was detected in gamma rays by \textit{Fermi}; it is listed as positionally associated with a source in the 4FGL-DR4 catalogue (\citealt{Abdollahi2020, Ballet2023}). A second object (GB6J1125+2610) is indicated only as a possible, low-confidence, association. Therefore, only a small fraction of the FSRQs that are likely associated with a neutrino event  seem to be clearly detected in gamma rays.
 Neutrinos produced via photo-pion interactions are expected to be accompanied by an equivalent
emission of $\gamma$-ray photons.
However, the neutrino production site is likely optically thick to high-energy photons \citep[e.g.][]{Boettcher2019}, and therefore, the gamma rays produced co-spatially with neutrinos are expected to be strongly absorbed and not detectable.
In this scenario, the observed $\gamma$-ray emission can be attributed to the leptonic component through the
inverse Compton mechanism. 
FSRQs with a low frequency peak ($<$10$^{13}$-10$^{14}$ Hz) usually have the high-energy hump peaking at energies $\leq$100 MeV, i.e. below the sensitivity of \textit{Fermi}. These blazars could appear weak in \textit{Fermi}, even if their luminosity at the peak is high.  
Notably, the detection of one (possibly two) sources is consistent with
the detection rate of bright FSRQs (flux $>$ 100 mJy) by \textit{Fermi}, which stands at approximately 40\%.

One of these objects (GB6J1125+2610, also known as TXS 1123+264) has been previously indicated as a possible neutrino emitter by \citet{Plavin2020}. Interestingly, a study of 20 years of archival very long-baseline interferometry (VLBI) data detected a significant brightening of the core after the neutrino event (\citealt{Komives2024}).  

Moreover, we compared the positions of the seven FSRQ candidates with the 9-year IceCube sky map \citep{Abbasi2022a}, which provides the probability that the observed neutrino flux is of non-astrophysical origin, mapped across the entire sky on a 0.2$^{\circ} \times$ 0.2$^{\circ}$ grid.
Interestingly, we find that four out of the seven FSRQ candidates lie within 1$^{\circ}$ of a significant peak (local p-value < 4$\times$10$^{-4}$) in the sky map
To calculate the number of associations expected by chance, we exploited the uniform distribution
of RQ quasar positions over the region by 
extracting 10,000 samples of seven positions each. We find a mean value of 0.12.
The most intriguing case is GB6J1350+233 (CRATESJ1350+23), located approximately 0.8$^{\circ}$ from the fifth most significant position on the IceCube scan \citep[Table S2]{Abbasi2022a}. Although the sky map and the IceCat-1 catalogue are not independent datasets, this correlation might suggest that, with a high probability, some of these seven FSRQs produce detectable and measurable neutrino fluxes.

\section{Bronze and gold events and the fraction of neutrinos produced by FSRQs}
In the IceCat-1 catalogue, neutrino events are classified as either bronze or gold, based on their estimated probability of being genuinely astrophysical in origin. Bronze events are assigned a 30\% probability, while gold events have a 50\% probability \citep{Abbasi2023a}. In the sample of 15 neutrino events considered in our analysis, six are flagged as gold and nine as bronze,
leading to an expected total of 5.7 real astrophysical neutrino events.
If we define an association as a positional match within 0.7$^{\circ}$, 8 out of the 15 neutrinos have at least one associated CLASH FSRQ, with one event (IC180125A) having two associations. Using our procedure to
generate control samples with randomized quasar positions, we estimate that only 1.5 such associations are expected by chance.
This implies 7–8 genuine associations, which is consistent with the expected 5.7 astrophysical,
at least under the hypothesis that a significant fraction of them are produced by FSRQs.
However, focusing only on the high-declination subsample yields a different picture. As noted, all seven neutrino events located at declinations above 20$^{\circ}$ are associated with a CLASH FSRQ, compared to only 0.8 associations expected by chance.
Among these, five events are bronze and two are gold, implying an expected 2.5 real astrophysical events, inconsistent with the observed 7 associations (probability $\lesssim$ 1\%).
As discussed earlier, the observed difference between the high- and low-declination samples may be due to unaccounted systematic uncertainties, affecting either the positional accuracy or the purity of the IceCat-1 catalogue. The fact that all seven high-declination events are associated with a nearby FSRQ suggests that this subsample has a higher astrophysical purity.

Assuming that all the 7 neutrino are real, this would imply that the fraction of neutrino produced by FSRQs is $>$60\% at 2$\sigma$ confidence.
Such a high fraction of high-energy neutrino events originating from FSRQs appears to be in tension with the much lower values reported in other studies  \citep[$<$20\%, ][and references therein]{Palladino2020, Oiko2022}.
The first argument in those studies is based on the expected electromagnetic counterpart, primarily in the gamma-ray band. These works typically estimated upper limits on the total neutrino flux from blazars by assuming a fixed  neutrino/$\gamma$-ray flux ratio. However, as discussed in the previous section, many FSRQs can appear faint or even undetectable by \textit{Fermi}, as their high-energy spectral peak can fall below the \textit{Fermi} LAT sensitivity range. As a result, the contribution of such sources to the neutrino flux could be significantly underestimated in gamma-ray-based analyses.

A second argument comes from the lack of significant neutrino point sources, which, combined with the observation of a diffuse astrophysical neutrino flux, implies that the dominant sources should form a dense population of weak emitters, rather than a few luminous ones.
In a recent analysis, \citet{Abbasi2023c} found that, assuming steady emission and  evolution following the star formation rate, a minimum local source density of  7$\times$10$^{-9}$ Mpc$^{-3}$ is required to match the observed neutrino flux. FSRQs, with a local density of only  5.6$\times$10$^{-11}$ Mpc$^{-3}$, as inferred from the \textit{Fermi} catalogue, are too rare to contribute significantly.
However, this conclusion is highly dependent on the assumed source evolution. In the same study \citep{Abbasi2023c}, the authors note that adopting the evolution of the AGN luminosity function would reduce the required source density by nearly two orders of magnitude, down to 6$\times$10$^{-11}$Mpc$^{-3}$ , much closer to the FSRQ density.

In our analysis we find that the neutrino-associated FSRQs have radio luminosities at 1.4 GHz in the range of 10$^{42}$-10$^{44}$ erg s$^{-1}$.  The space density and redshift evolution of this population have been well characterized in previous works \citep{Mao2017, Caccianiga2019, Diana2022}. Their local space density is estimated to be around 4$\times$10$^{-10}$ Mpc$^{-3}$, with a strong positive evolution peaking at z $\sim$ 1.9, where the density reaches 2.5 $\times$10$^{-8}$ Mpc$^{-3}$. This evolution is significantly faster than that of the star formation rate and more closely resembles that of AGNs. Taking this into account, our findings are consistent with both the absence of neutrino multiplets and the observed diffuse neutrino flux and thus reconcile the high apparent association rate of FSRQs with the broader neutrino source population. A similar result has been found by \citet{Neronov2020}.

\cite{Abbasi2022c}, using a stacking analysis of 137 FSRQs selected in the 30–100 MeV band,
found that their contribution to the unresolved background is $\lesssim 1\%$.
At first glance, this result may seem to be in tension with our findings, since some level of soft $\gamma$-ray/hard X-ray emission is inevitably expected from FSRQs if they are indeed neutrino emitters (see the discussion in \cite{Abbasi2022c}. However, the expected contribution of these very bright and rare objects to the unresolved neutrino flux is not significant. Furthermore, the hadronic contribution to the soft $\gamma$-ray flux is difficult to quantify, while the inverse Compton emission is certainly contributing in this band. It is therefore plausible that the 137 FSRQs bright in the soft $\gamma$-ray band considered by \cite{Abbasi2022c} are simply those with the strongest Compton dominance, and thus not necessarily the most promising neutrino sources. We do expect that the FSRQs discussed in this work emit in the soft $\gamma$-ray band, but it is likely that the sensitivity of \textit{Fermi} in this energy range is insufficient to detect them.

\begin{table*}
\scriptsize 
 \centering
 \caption{Possible neutrino sources.}
 \label{tab:matches1}
 \begin{tabular}{l l c c c c c c c c c c c}
  \hline
Quasar Name & Other Names & RA & Dec.       & z & S$_{5GHz}$  & Log L$_{bol}$  &Log$M_{BH}$  & Edd. ratio & \textit{Fermi} & IceCat-1& dist\\ 
         &             & (deg)& (deg)   &   & (mJy)       & (erg s$^{-1}$)  &(M$_{\rm\sun}$) & &  & & (deg) \\  
\hline
GB6J0808+495 & 5BZQJ0808+4950  &  122.1652  & 49.84348  &  1.44 &   1315  &  46.58  &    8.84  &    0.43 & Y &    IC110929A & 0.50 \\
GB6J1126+451 &  5BZQJ1126+4516  &  171.7402  & 45.26843  &  1.81 &    360    &  47.33  &    9.42  &    0.65 & N &    IC111012A & 0.63 \\
GB6J1125+261 &  5BZQJ1125+2610  & 171.4737  &  26.17222  & 2.35  & 1176    & 47.32 &     9.29   &   0.84  & Y ? &    IC120523A  & 0.44 \\
GB6J0853+281 & 5BZQJ0853+2813  &  133.3242  & 28.23054  &  0.92 &    109  &  46.45  &    8.83  &    0.33 &  N&    IC150904A & 0.42 \\
GB6J1350+233 & CRATESJ1350+23  &  207.6902  & 23.52922  &  0.53 &    141  &  45.91  &    8.77  &    0.11 & N &    IC180125A & 0.29 \\
GB6J1623+390 & 5BZQJ1623+3909  &  245.7817  & 39.15900  &  1.98 &    244  &  47.67  &    9.66  &    0.80 & N &    IC190201A & 0.66 \\
GB6J1027+480 & 5BZQJ1027+4803  &  156.8045  & 48.05376  &  1.29 &    242  &  46.27  &    9.16  &    0.10 & N &    IC200806A & 0.42 \\
\hline
\end{tabular}
\end{table*}

\section{Discussion}
Several studies aimed at the search for positional correlations between IceCube events and extragalactic sources were carried out in the last few years. They all adopted different approaches and started from different neutrino events and AGN catalogues. 
In the following sections we summarize the main results coming from the studies. 
\subsection{Neutrino events}
Concerning neutrino events, many authors (including us) considered the list of track-like events with relatively good positional accuracies (\citealt{Giommi2020a, Plavin2020, Plavin2023}).Early searches used cascade-type events (\citealt{Padovani2016}), which are characterized by larger positional uncertainties but have a more likely astrophysical origin. Apart from the list of single neutrino events, the IceCube collaboration has also provided an all-sky map of possible neutrino excesses (\citealt{Aartsen2017}) in the form of a p-value, for each sky position, that gives the level of clustering in the neutrino data, i.e. a measure of the significance of neutrino events being uniformly distributed. The map is based on more than 700,000 events taken during 7 years of data collection. This map was used by the IceCube collaboration (\citealt{Aartsen2017}) and by other authors (\citealt{Plavin2021, Buson2023, Bellenghi2023}) to search for astrophysical counterparts.
More recent versions of the map, based on 10 years of data, were produced and used by \citet{Aartsen2020} and \citet{Bellenghi2023}.

\subsection{FSRQs versus BL Lacs}
As far as the blazar catalogues are concerned, different possibilities were considered. A common choice was the selection of gamma-ray blazars, detected in one of the available \textit{Fermi} catalogues (\citealt{Aartsen2017, Aartsen2020}) or blazars (typically BL Lacs) that are promising candidates to be detected at very high-energy gamma rays ({\citealt{Giommi2020b}). This choice was motivated by the fact that, in the hadronic model, a neutrino emission should be accompanied by the emission of a gamma-ray photon. However, not only the gamma-ray photon can be absorbed, right after its emission or during its travel towards Earth (by the  diffuse extragalactic background light), but a gamma-ray emission can be also produced by other mechanisms (e.g. the inverse Compton scattering) without the production of a neutrino (leptonic model). It is possible that both leptonic and hadronic mechanisms are at work in the jets of blazars, and therefore, a selection based on gamma-ray strength  may not be the optimal choice.
A gamma-ray `agnostic' approach was adopted by \citet{Buson2022,Buson2023} and \citet{Plavin2020,Plavin2021, Plavin2023} who considered, respectively, all the blazars included in the Roma-BZCat catalogue (5BZCat; \citealt{Massaro2015}) and the radio brightest blazars, based on the Very Long Baseline Array calibrator surveys. In all these works, blazars are analysed as a unique class, without distinguishing between BL Lac objects and FSRQs.

Overall, the most significant ($\geq$3~$\sigma$ post-trial) correlations found so far are those with the BZCat blazars (\citealt{Buson2022}) and with radio bright ($>$150~mJy, \citealt{Plavin2020, Plavin2021}) blazars. In all these works, the FSRQs that seem to correlate with the neutrino events are radio bright, above $\sim$200 mJy. In the `dissection' analysis discussed in \citet{Giommi2020a}, a $>$3$\sigma$ (post-trials) correlation is found only with blazars (including both BL Lacs and FSRQs) with the synchrotron component peaking at high frequencies ($>$10$^{14}$~Hz) while no significant correlation is found with low-frequency peaked blazars. 
\textit{Fermi}-bright blazars do not seem to provide significant correlations except when using the most recent (10~years) map (\citealt{Aartsen2020}). The correlation with radio bright blazars from the RFC catalogue \citep{Petrov2025} was confirmed at 3.3$\sigma$ also in the southern hemisphere by \citet{Bellenghi2023} using the 7-year map although nothing significant was found using the revised map based on 10 years of data. This could be in part due to the lower sensitivity achieved by their analysis compared to \citet{Aartsen2020} caused by the lack of an accurate  detector response matrices publicly available (\citealt{Bellenghi2023}).

In the work presented here, we obtain a significant correlation  ($\sim$4$\sigma$) based on a sample of radio bright ($>$100~mJy) FSRQs with declinations above 20$^{\circ}$. The type of selection is similar to that discussed in \citet{Plavin2020} although we focused only on the FSRQs, excluding BL Lac objects. Moreover, we restricted the analysis to the sky area covered by SDSS in order to have an almost complete and uniform spectral identification of the radio bright sources. It is interesting to note that the four sources in the \citet{Plavin2020} that are indicated as most likely counterparts of the neutrino events are all classified as FSRQs, even if the starting sample includes both FSRQs and BL Lac objects.
These four FSRQs are not included in our sample since they fall in an area of sky that we are not covering.
Our result is corroborated by the observation that at high declination (Dec. $>$20$^\circ$) the significance of the correlation is higher.
Based on these results, it seems that the radio-brightest FSRQs, mostly of low-frequency peak type, give a high signal in the correlation analyses carried so far. 

We note that the `dissection' analysis performed by \citet{Giommi2020a} did not find any significant correlation when considering low-frequency peaked blazars. This could be due to the fact that the objects considered in that work include both BL Lacs and FSRQs and, more importantly, radio faint sources. For a direct comparison with our findings, this kind of analysis should only be carried out on bright FSRQs ($>$100~mJy).

We finally note that TXS0506+056 is formally classified as a BL Lac object and, therefore, a priori excluded in the analysis presented here. The fact that broad emission lines, although present (\citealt{Padovani2019}), were not visible in the discovery spectra is related to the fact that the synchrotron component peaks at a relatively high frequency peak (10$^{14}$-10$^{15}$ Hz, \citealt{Padovani2018}), i.e. in the visible range, which is higher compared to `normal' FSRQs, where this component typically peaks in the far infra-red band. Because of this peculiar characteristic, the non-thermal emission from the jet is able to swamp the accretion disk and BLR components, thus hiding the emission lines. These objects, also called `masquerading BL Lacs' (\citealt{Giommi2013}), represent a rare tail of the entire FSRQ population. Distinguishing these sources from `true' BL Lacs is not simple, but their inclusion in a future analysis might increase the significance of our results.  

\subsection{Jet versus accretion}
As mentioned in the Introduction, neutrino production mechanisms linked to matter accretion onto the SMBH have been proposed in the literature for both NGC 1068 and TXS 0506+056. However, the results presented in this paper suggest that the neutrino production mechanism in these FSRQs is not associated with accretion. The correlation between FSRQ positions and neutrino events, in contrast with the lack of correlation of samples of quasars with similar optical properties but no radio emission, strongly indicates that neutrino emission originates in relativistic jets.
To our knowledge, it is the first time that such a direct comparison between radio-loud and RQ quasars, in the search for counterparts of neutrino events, is carried out. 

We note that our result is based solely on the positional correlation between neutrino events and radio jet emission. Other studies have found evidence of temporal correlations between radio activity and neutrino detections \citep{Plavin2020}.
Since the radio flux densities used in our analysis were measured at epochs completely independent of the neutrino events, our approach does not allow us to test for a direct temporal connection between radio emission and neutrino production.
In the lepto-hadronic scenario, which appears to provide the best match to observational data, the radio emission is primarily produced by synchrotron radiation from the leptonic component \citep[e.g.][]{Petropoulou2020} and is not directly associated with the $p\gamma$ interactions responsible for neutrino generation.
The high radio flux levels observed in our analysis can be interpreted as a selection effect, with neutrinos being more likely detected from jets that are either intrinsically more powerful or strongly Doppler boosted \citep{Plavin2023}.

\section{Summary and conclusions}
Starting from an optically selected sample of AGNs \citep{Shen2011} and a well-defined, flux-density-limited radio survey at 5 GHz in the northern hemisphere (CLASS), we identified a sample of 1122 FSRQs, referred to as the CLASH sample, from the larger RQ quasar population with identical optical properties.
We then performed a simple positional cross-correlation between both samples and a set of 15 IceCube events extracted from IceCat-1, selecting events with relatively well-constrained positions within the SDSS/CLASS survey area. Using a Monte Carlo randomization of positions, we estimated the p-values of the observed correlations.
Our analysis of the CLASH sample at six different radio flux density limits reveals that the most significant correlation occurs at a flux density limit of 100 mJy, corresponding to a post-trial significance of 2.7$\sigma$.
In contrast, applying the same test to the RQ quasar population yielded significantly higher p-values that were fully consistent with a random distribution.

Analysing the distribution of minimum distances between neutrino events and CLASH sources, we observe a clear trend: at declinations above 20$^\circ$, these distances are systematically smaller than what would be expected by chance, whereas closer to the celestial equator, they tend to be larger.
A KS test confirms this trend. While the minimum distance distribution for the full sample differs from a random distribution at a moderate significance level of 2.2$\sigma$, restricting the analysis to high-declination events increases the statistical significance to 4.0$\sigma$.
In contrast, we find that the minimum distance distribution for the RQ quasar sample is indistinguishable from a random sample. This  reinforces the observed correlation between neutrinos and FSRQs.

These findings indicate that a majority ($>$60\%) of the high-energy astrophysical neutrinos detected by IceCube originate from blazars, particularly FSRQs. Their production likely occurs in relativistic jets rather than in the RQ regions, such as the accretion disk or corona.
Our expectation is that most neutrino events that will be detected at high declinations by IceCube, particularly those with relatively good positional accuracies, will be located close to a luminous source with a flat radio spectrum and broad emission lines in the optical.

\begin{acknowledgements}
The authors thank Paolo Giommi and Lorenzo Caccianiga for valuable discussions and suggestions. 
We acknowledge the financial support from INAF under the projects “Quasar jets in the early Universe” (Ricerca Fondamentale 2022) and “Testing the obscuration in the early Universe” (Ricerca Fondamentale 2023).

\end{acknowledgements}

\bibliography{aa54527-25}{}
\bibliographystyle{aa}



\end{document}